# Ultrastrong, Ultraflexible, and Ultratransparent Polyethylene Cellular Nanofilms

Ping GAO[1*], Qiao GU[1†], Jin LI[1†], Runlai LI[1], Qinghua ZHANG[1], Lu-tao WENG[1,2], Tianshou ZHAO[3], T.X. YU[3], Minhua SHAO[1], and Khalil AMINE[4*]

[1]*Department of Chemical and Biological Engineering, The Hong Kong University of Science and Technology, Clear Water Bay, Kowloon, Hong Kong SAR, PR China.*

[2]*Materials Preparation and Characterization Facility, The Hong Kong University of Science and Technology, Clear Water Bay, Kowloon, Hong Kong SAR, PR China.*

[3]*Department of Mechanical and Aerospace Engineering, The Hong Kong University of Science and Technology, Clear Water Bay, Kowloon, Hong Kong SAR, PR China.*

[4]*Chemical Science and Engineering Division, Argonne National Laboratory, Argonne IL 60439, USA*

† *These authors contributed equally to this work.*

*Corresponding author. E-mail: kepgao@ust.hk (Ping GAO); co-corresponding author. Email: amine@anl.gov (Khalil AMINE).

## ONE-SENTENCE SUMMARY

We created an ultrastrong, ultraflexible, and ultratransparent polyethylene cellular nanofilm consisting of Delaunay cells with highly crystalline anisotropic cell edges, and the freestanding nanofilm was fabricated into a transparent respiratory face covering.

**ABSTRACT**

Light weight and mechanically robust cellular polymer nanofilms provide materials solutions to many cutting-edge technologies, such as high-flux membrane filtration, ultrathin flexible energy storage, and skin-conformable devices. However, it remains challenging to fabricate hand manipulatable cellular polymer nanofilms for use as self-standing structural materials. Herein, we used a sequential planar extension approach to transform low-entanglement ultrahigh molecular weight polyethylene (UHMWPE) gel films dispersed in porogenic polyethylene oligomers into cellular nanofilms consisting of stretch-dominated triangular cells of molecularly anisotropic cell edges. The microstructure afforded the cellular nanofilm, which had a thickness down to 20 nm, with a unique combination of ultratransparency (>98.5%), ultrahigh in-plane tensile strength (1071 MPa cm$^3$g$^{-1}$), and ultrahigh flexibility: a 43 nm thick film can deflect reversibly up to 8.0 mm in depth (185,000 times) under a spherical indentation load. As an application, we fabricated the nanofilm into a freestanding ultratransparent respiratory face covering. The new polyethylene cellular nanofilms are expected to represent a new class of platform membranes for advancing fundamental and technological development.

**MAIN TEXT**

The Covid-19 pandemic has spurred a worldwide demand for face coverings, but opaque face masks currently in use are not suitable for the recognition of faces or speech. Thus, the need for face mask removal at security checkpoints may lead to exposure risks; and people with impaired hearing difficulties are unable to communicate through body language, such as by lip reading. Breathable ultralight cellular dielectric polymer nanofilms with thicknesses on the order of 100 nm may present a potential solution for transparent face masks; better still, if the cellular polymers have a large bandgap, they will have the added advantage of being disinfected with ultraviolet (UV) light. The key limiting factor for the use of cellular polymer nanofilms is their mechanical properties (*1–3*). Indeed, we often discuss thin film-based devices in terms of their optoelectronic properties, and many such devices are limited by their mechanical properties. Even nanofilms supported on porous substrates for reverse osmosis need to withstand fouling-induced friction during membrane filtration (*4, 5*).

The mechanical strengths of cellular polymer nanofilms depend both on their topology and the polymers. Topologically, an open-cell polymer nanofilm may be considered a randomly interconnected network of nanostructures that form the edges of the cells (*6*). The mechanical strengths of a cellular film follow the porosity scaling equation: $\sigma_p \propto \sigma_s (1-\varepsilon)^n$, where $\sigma_p$ and $\sigma_s$ are the mechanical strengths of the porous and the solid polymer, respectively; $\varepsilon$ is the volumetric porosity, $\varepsilon = 1 - \frac{\rho_p}{\rho_s}$, where $\rho_p$ and $\rho_s$ are the density of the porous and solid polymer, respectively; and the exponent $n$ in the scaling relation is determined by the pore edge connectivity. The value of $n = 1$ for highly connected stretch-dominated Delaunay cells, and $n > 1.5$ for bending-dominated Voronoi tessellations (*6*). Topologically, cellular polymer nanofilms with thicknesses less than 100 nm may be considered 2-D films. The metric $M$ for Maxwell stability (6) in 2D is $M = b - 2j + 3$, where $b$ is the total number of struts (edges) and $j$ is the total number of frictionless joints. For stretch-dominated lattice cells, $M > 0$ (6). Triangulated Delaunay lattices, which are the smallest polygons in 2D, possess the highest cell edge connectivity ($b$) and thus provide cellular films with the highest stretch-dominated stability. The stretch-dominated Delaunay triangulations enable the principal mode of deformation by extension/compression; therefore, their failure mode is by buckling rather than by bending. However, cellular polymer nanofilms reported hitherto (to our knowledge) are mostly formed by homogeneous phase separation processes characterized by bending-dominated Voronoi tessellations (7), and the mechanical properties of cellular films with Voronoi tessellations deteriorate rapidly with porosity. The property of the parent polymer, $\sigma_s$, is the other critical factor that determines the ultimate mechanical properties of a porous film. One of the literature strategies for preparing strong polymer nanofilms has been through the use of reinforcing fillers or intrinsically strong polymers. As a result, solid silk fibroin nanofilms that were reinforced by crystalline β-sheets were reported to have tensile strengths up to 100 MPa; however, the nanofilms were extremely brittle (ductility less than 0.5%) due to confinement-induced low crystallinity (8). High-strength and high-stiffness gold nanoparticle-reinforced polymer nanofilms prepared by layer-by-layer deposition were also reported. The modulus and strain to failure estimated using pressurized bulging tests that the film had a stiffness of 40 GPa and maximum ductility of 2%, although the measurements were conducted using indentation measurements on small samples (the maximum displacement was ~50 $\mu m$ for a freestanding film over a span of 600 μm)(9). Recently, conductive freestanding parylene nanofilm (~100 nm) reinforced with silver nanowires was found to be strong enough to function as a skin conformable speaker (*10*); however, its non-porous structure may hinder skin perspiration when used for practical applications. In addition, the difficulty with handling fragile cellular nanofilms has made direct in-plane mechanical property measurements challenging (*11*); most measured tensile properties are limited to indentation measurements where large sample areas are used to maximize measured load signals (*12*, *13*).

Ultrahigh molecular weight polyethylene (UHMWPE, with weight averaged molecular weight $\overline{M}_W \geq 10^6$ kg kmol$^{-1}$) possesses one of the highest theoretical modulus values and tensile strengths among synthetic polymers ($E_s \sim$216-360 GPa and $\sigma_s \sim$33-66 GPa) in its fully extended all-planar zig-zag (diamond-like) single-crystal orthorhombic structure (*14*), which makes it one of the strongest and lightest synthetic materials used in the fabrication of bullet-proof vests (*15*, *16*). It is also one of the best insulating polymers due to its high band gap (~8 eV) (*17*). Therefore, it is an ideal material to transform into robust transparent cellular nanofilms.

Herein, we report a new and scalable two-step approach to design ultrastrong, ultraflexible, and ultratransparent cellular polyethylene nanofilms. Our approach involves sequential planar extensions whereby low-entanglement thin gel films (~200 $\mu m$) of UHMWPE consisting of folded-chain lamellar crystals dispersed in 95% polyethylene oligomers are first transformed into extended-chain fibril crystals and then weaved into 2D triangulated Delaunay cellular lattices. The topological and conformational microstructures endow the newly developed polyethylene nanofilms (down to 20 nm in thickness) with a unique combination of mechanical, optical, and porous properties. A prototype air respirator made using the freestanding ultrathin ultratransparent cellular film with an average cell diameter of ~100 nm showed an air flow rate of 85 L/min at a transmembrane pressure drop of 146 Pa and an aerosol particle filtration ($75 \pm 20\ nm$) efficiency of 99.8%, satisfying the standard requirement for use as a transparent air respirator (*18*). This work demonstrates that our new approach to geometrical and molecular conformation manipulations can provide highly functional polymer nanofilms that can enable a broad range of fundamental and technological breakthroughs.

**Method of Polyethylene Nanofilm Preparation**

The novelty of our approach for the transformation of low-entanglement gel crystallized films lies in the use of sequential biaxial planar extensional deformations in conjunction with the choice of polyethylene oligomers as the porogenic molecular lubricant. The polyethylene oligomers disentangled chains in the molten state, lubricated chains unravelling from lamellar crystals into fibril crystals, eased fibril splitting and rotation for network formation, and created sites for pores after being extracted by a solvent (*19*). The use of sequential biaxial planar deformations for the processing of the low-entanglement gel-crystallized thin film method was chosen because it provided networks of oriented polymer fibre crystals *in situ*, ensuring high mechanical integrity and high crystallinity of the polymer. This simple, reproducible and easily controllable technique has been used for the processing of safe battery separators (*20*) and high-flux water distillation membranes (*21*).

**Figure 1, A and B** show a schematic of the two-step approach to transforming UHMWPE thin precursor gel films. In our approach, we chose to use a top-down sequential biaxial planar extension method followed by solvent extractions under lateral

constraints. As illustrated in **Fig. 1A**, the low-entanglement gel precursor of semi-crystalline UHMWPE (folded-chain crystals) dispersed in polyethylene oligomer was stretched sequentially under planar extension in the plane of the film at temperatures slightly lower than the melting temperature of UHMWPE (*20*) (see **Supplementary Materials**). Lateral contractions were prevented by applying a lateral constraint plus the virtue of the thinness of the initial gel film lubricated by the molten polyethylene oligomers, which also served to maintain incompressibility. The planar extension of the film may be characterized by the Cauchy-Green deformation tensor $C$ for planar extension of incompressible materials and is given by (*22*): $C = \begin{pmatrix} \lambda^2 & 0 & 0 \\ 0 & 1 & 0 \\ 0 & 0 & 1/\lambda^2 \end{pmatrix}$, where $\lambda$ is the extension ratio along the extension axis $x_1$. The planar deformation is thus characterized by the equality of the film extension ratios along the length of the film to that of the film contraction ratios in the thickness of the film whilst maintaining the width of the film constant. This is illustrated in **fig. S1B**, where no apparent width contraction of the film was discernible during planar extension. Characterization of the film thickness (to be shown in the thickness characterization section) showed that the deformation in the film followed the Cauchy-Green deformation tensor (C) completely throughout the sequential extension process, i.e., the film thickness was inversely proportional to the total film extension ratios. A gel film with an initial thickness of ~200 $\mu m$ became approximately 400 nm thick after it was stretched sequentially by a total extension ratio of $\lambda = 22 \times 22$. The final film thickness was 21.7 nm for a solid film after solvent extraction of the polyethylene oligomer (95 wt%) under lateral constraints.

Microscopically, the initially randomly oriented lamellar crystals (platelets in **Fig. 1A**) in the gel film under the extension stresses became predominately aligned along the $x_1$ axis. In the meantime, the stress acting along $x_2$ axis causes crystal alignment along the $x_2$ axis. At sufficiently high extension ratios, chains in the lamellar crystals became pulled out of these lamellar crystals and formed extended-chain fibril crystals (*23*). The carbons along the chains in extended-chain fibril crystals form zig-zag conformations like those zig-zag carbons in a diamond (**Fig. 1B and fig. S1A**), giving the film strong mechanical property along the fibril axis (*24*). During the subsequent planar extension in the orthogonal direction ($x_2$) or along the $x'_1$ − axis, in addition to further chain alignment (predominantly along the $x'_1$ axis now), topological transformation into triangulated networks took place. The fibrils oriented perpendicular to the extension stress became split (**Fig. 2D**), rotated (assisted by trans-fibrils) and bent simultaneously to form stretch-dominated networks (we believe). Weak interfibrillar forces or a low shear modulus led to fibril splitting, which was enhanced by the trans-fibrils formed along the $x_2$ axis during the first planar extension process due to planar extension stress in the transverse axis of the extension direction. A thickness contraction strengthened fibril interlocking by increasing the number of fibril contacts, forming pinned joints among the fibrils and leading to stretch-dominated triangulated lattice cells. Polygons with low edge connectivity, e.g., quadrilaterals or hexagons, were unable to resist a stretching load in the plane or along the diagonal of the quadrilaterals and therefore were unable to withstand the planar extensional stresses acting in the plane of the film (*6*). To simplify the microstructure in a cellular nanofilm, we introduced a thermal treatment by annealing the film at temperatures close to the peak melting temperature of the UHMWPE while holding the lateral dimensions constant so that fibres at the contact points across the film thickness fused to form a monolayer 2D cellular film,

**Fig. 1B (Fig. 2, C to F)**. The thermally induced shrinkage in the plane of the film caused tension in the plane of the film, and thus, the fibrils in the film responded by harnessing nanofibres from different planes into a triangulated Delaunay structure – **see Supplementary Materials, fig. S1B**. In other words, the topological construction of the lattice structure (stretch-dominated open cell structure) formed in the $2^{nd}$ planar extension step was simply driven by system energy minimization.

The combined topological and extended-chain conformation (**Fig. 1B**) plus the nanometer thickness in the newly developed polyethylene nanofilm provided the nanofilms with exceptionally high mechanical strengths and optical transparency; we refer to our new nanofilm as GP Nano. As shown in **Fig. 1D**, GP Nano exhibited the highest specific tensile strength among a range of high-performance materials. Its specific strength was $1071 \pm 75$ MPa g$^{-1}$ cm$^3$, which is approximately 17 times that of SS304 stainless steel and three times that of a densified wood (*25*). The GP Nano was also ultratransparent, showing optical transmittances greater than 98.5% in the visible and infrared regions (**Fig. 2A**), as depicted by a photograph of the film confined inside a carbon fibre frame (**Fig. 1C**) that was taken at the outdoor sportsground of HKUST. Further quantitative mechanical data are presented in **Fig. 3**. To put this simply, a 15 mm wide GP Nano with a thickness of 100 nm lifted 200 grams of water, which is approximately four times that expected from a single-crystal monolayer of graphene with the same width (*26*). However, no hand manipulatable atomically thin single-crystal graphene is yet available; to the best of our knowledge, GP Nano is the first hand manipulatable cellular nanomembrane.

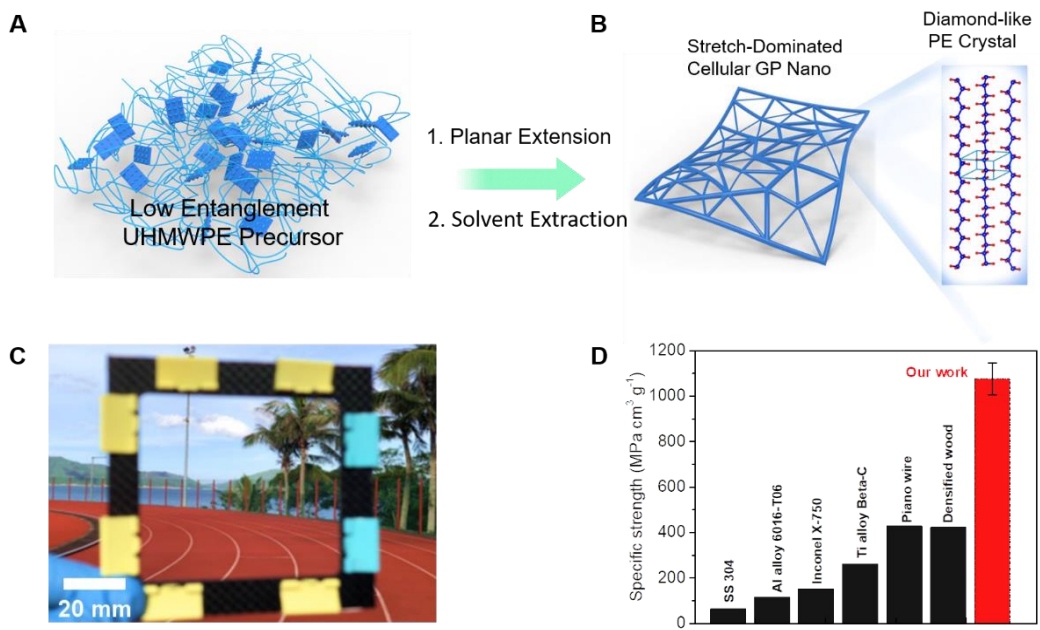

*Fig. 1. Processing approach and optical and mechanical performance of cellular ultrahigh molecular weight polyethylene (UHMWPE) nanofilm. (A) A schematic of the low-entanglement semi-crystalline UHMWPE (folded-chain crystals damasked in bright blue) gel film containing 5 wt% UHMWPE in polyethylene oligomer. (B) A schematic of the structure of the 2D cellular nanofilm consisting of topologically stretch-dominated triangulated cells with the highly crystalline cell edges in which the polyethylene chains take zig-zag conformations like the carbon chains in a diamond crystal. The extended chain illustration was produced using the VESTA software (27). (C) Photo of a freestanding cellular nanofilm contained inside a carbon fiber frame taken at the outdoor sportsground of HKUST campus. (D) The film exhibits in-plane specific tensile strength (1071 ± 75 MPa cm³ g⁻¹) which is approximately 3 times that of densified wood (25), (28–32).*

**Thickness and Microstructural Characterization**

To take advantage of the large areas of GP Nano, we used a simple weighing method to determine its thickness (**see Supplementary Materials**). Briefly, the weights of samples with film areas about 45 cm$^2$ were measured on a Sartorius M500P ultramicrobalance, and before they were weighed, the optical transmittance spectra of the samples were collected to correlate the sample thickness with the optical transmittance. Ultralight cellular GP Nano films with area densities ranging from 0.0217 g/m$^2$ to 0.0986 g/m$^2$ were successfully fabricated. These area densities correspond to nanofilms with solid thicknesses from 21.7 nm to 98.6 nm, assuming a mass density of polyethylene of 1 g/cm³ (*24*). All nanofilms were ultratransparent, as shown in **Fig. 2A**. The thinnest nanofilm (21.7 nm) had an optical transmittance greater than 98.5% at wavelengths longer than 400 nm and 88.5% at far UV irradiation of 200 nm. The optical transmittance was even higher than that through a monolayer graphene (*33*), which shows 97.7% optical transmittance at wavelengths beyond 550 nm. The high optical transparency in the far UV region was expected because of the large polyethylene bandgap of 8 eV(*17*), which renders optical attenuation by light scattering only when the incident irradiation wavelengths are longer than 200 nm. We consequently modelled the optical attenuation against mass area density, $\omega_s$ (µg/$cm^2$), using the Beer-Lambert law (*34*) (**Fig. 2B**):

$$log_{10}\left(\frac{T_0}{T_{200}}\right) = 0.0246\ \omega_s \qquad (1)$$

where $T_0$ and $T_{200}$ are the optical transmittance through air and through the GP Nano at a monochromatic irradiation of 200 nm, respectively, and the constant 0.0246 is the fitting constant by least square regression analysis. The optical attenuation of the film followed **Equation (1)** closely at area densities up to 8 µg/$cm^2$ (80 nm solid film thickness), and the thinner the film was, the smaller the error bars and the better the agreement between the experiments and the model prediction shown by **Equation (1)**. This further confirmed our hypothesis that the optical attenuation through GP Nano was mainly caused by the surface scattering effect since any light absorption would have rendered nonlinearity in the optical attenuation. Moreover, in the subsequent analysis, we simply employed **Equation (1)** for the calculation of GP Nano's thicknesses.

The micrographs of GP Nano, both prior to and after thermal annealing (partial fusion) treatment, show the presence of interconnected stretch-dominated triangulated cells.

The SEM micrograph of the GP Nano prior to thermal annealing is depicted in **Fig. 2C**, in which convoluted triangulated cells were observed to reside across different layers through the film thickness. TEM characterization of the structure, **Fig. 2D**, shows that the triangulated cells were formed by overlapping fibrils from different layers. It is also noted that these fibrils may be categorized into two different groups based on the width: thick fibrils had widths of $18.8 \pm 5.3\ nm$, and thin fibrils had widths of $4.6 \pm 1.9\ nm$; the thick fibrils split to form the thin fibrils (see inset, **Fig. 2D)**. Image analysis of the micrograph depicted in **Fig. 2C** using Image J showed that the cells had pores with diameters of $106 \pm 69\ nm$. Constrained thermal annealing transformed the layered structure into a monolayer 2D triangulated cellular topology. Furthermore, this was also accompanied by a significant increase in cell diameters, cell edge widths, and pinned-joints (**Fig. 2, E and F)**. The average pore size become approximately three times that of the unannealed film ($0.35 \pm 0.27\ \mu m$). The porosity in the film estimated by image analysis of the SEM micrograph in **Supplementary Materials**, **fig. S4** with Image J was 54.15%.

Having established the triangulated lattice cell topology of the film, we probed the chain conformations of the cell edges in GP Nano using wide angle X-ray diffraction (WAXD) and differential scanning calorimetry (DSC). The polycrystalline Debye-Scherrer rings depicted in the WAXD pattern **(Fig. 2G)** show that the average chain alignments were homeotropic in the plane of the film. The eight diffraction peaks from the polyethylene orthorhombic unit cell, including the strong (110), (200), and (020) peaks (except the weak (120) peak)(*35*), indicated that the crystallites were large in the GP Nano. By fitting the diffraction peaks with Gaussian function, as shown in **Fig. 2H**, the apparent crystallite sizes perpendicular to the orthorhombic (110) and (200), planes were calculated to be 14.29 nm, 9.48 nm (*36*), respectively. The corresponding crystallite sizes were 13.80 nm and 12.48 nm after the first planar extension (see **fig. S5 & table S1**). Thus, the crystallites mainly decreased in size normal to the (200) plane during the second planar extension process, suggesting that a reduction in the film thickness occurred along the a-axis of the crystals.

DSC of the GP Nano showed that approximately 30 % of the crystals in the films were extended-chain crystals (**Fig. 2I**). This was estimated by taking the ratio of the heat absorbed between 145.5 °C and 167.3 °C, which are above the equilibrium melting temperature of the orthorhombic crystals in polyethylene (145.5 °C ) (*37*) over the total heat absorbed in the first heating trace (**fig. S6**). In order to further confirm this, we thermally annealed the film sample at 145 °C in a carbon fibre frame before its characterization under DSC. The measured data on the annealed film (**Fig. 2I**) depicted two clearly resolved melting peaks: (a) the low-temperature melting peak ($T_{m1}$ = 130.0 °C) corresponded to the epitaxially grown melt-crystallized folded chain crystals, and (b) the high-temperature melting peak ($T_{m2}$ = 150.0 °C) corresponded to that of the extended-chain fibre crystals (*38*).

In addition to consisting of extended-chain fibre crystals with large crystallite sizes, the GP Nano also exhibits a high crystallinity of approximately 75.5%; the crystallinity was estimated using the melting endotherm of the first heating trace in **Fig. 2I** (see **fig. S6**). Upon complete melting, its crystallinity decreases to 48.5% (**Fig. 2I**, Control reheat), which corresponds to the typical values of for melt crystallized high-density polyethylene. The high crystallinity and high molecular orientation of the GP Nano demonstrated the merit of our approach to nanofilm fabrication, which provided the GP

Nano cell edges with high mechanical strengths. In addition, reports have shown that low crystallinity in polymer nanofilms is an important cause of brittleness in nanofilms (*8*); however, it is not easy to circumvent the nanoconfinement-induced low crystallinity with the existing approaches for nanofilm synthesis.

The meta-structure of GP Nano that consists of triangulated lattice cells with cell edges composed of extended-chain fibre crystals was expected from the scaling law to exhibit strong mechanical performances (*6*), as is presented in the next section.

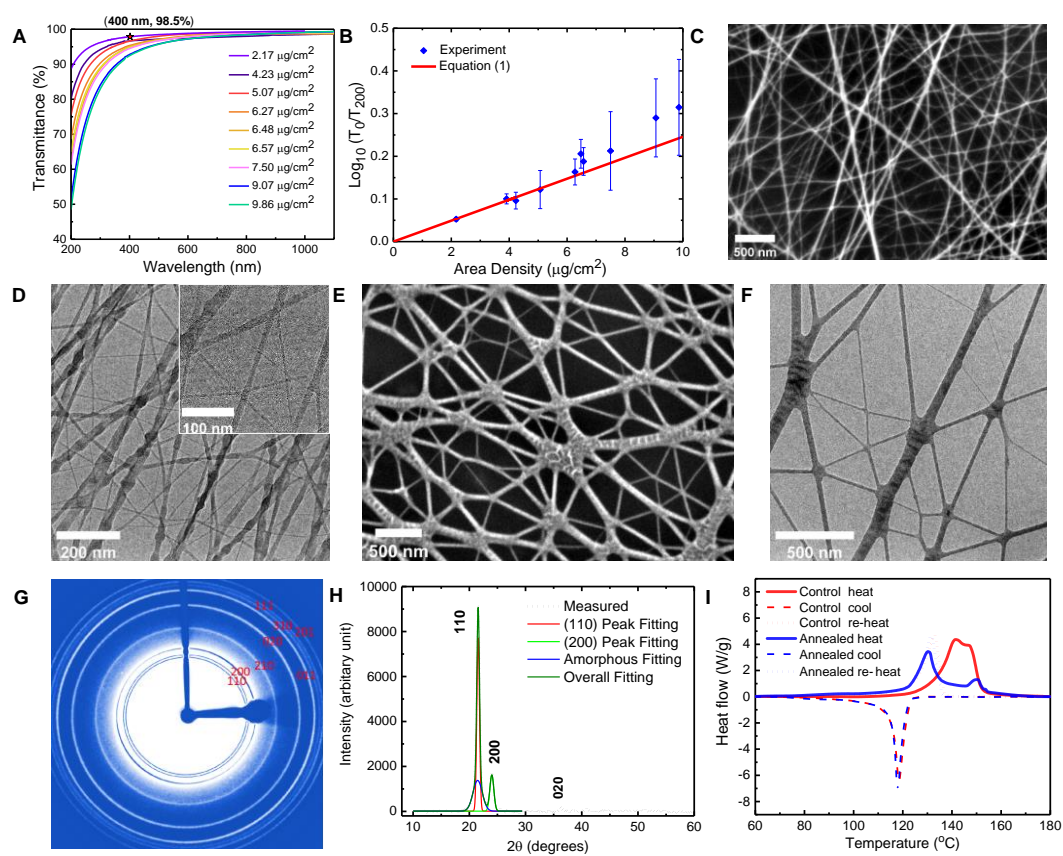

*Fig. 2. Thickness and microstructure characterizations of the GP Nano (Cellular UHMWPE Nanofilms).* **(A)** *UV-Vis spectra of GP Nano versus GP Nano's mass area density. The purple star label here denotes the optical transmittance @400 nm of the lightest sample with area density 0.0217 g/m². denoting optical transmittance 98.5%.* **(B)** *Plots of optical attenuation versus GP Nano's area density together with the fitting with the Beer-Lambert's law. The scattered points are measured optical attenuation together with the error bars caused by mass measurements; the continuous line represents the least square fitting to the Beer-Lambert law.* **(C), (D)** *SEM and TEM micrographs of GP Nano with solid film thickness 21.7 nm taken on a free-standing sample suspended on a copper grid without any*

*conductive coating (see **fig. S3**). The inset in **Fig. 2D** is a larger magnification of the structure depicting triangulated topological cells and cell edges splits from thicker fibrils. **(E), (F)** SEM and TEM micrographs of GP Nano after constrained thermal annealing at 140 ℃. The film sample was with area density 0.0217 g/m², the micrograph was taken on a free-standing sample suspended on a copper grid without any conductive coating (see **fig. S3**). **(G)** 2D Debye-Scherrer WAXD patterns of GP Nano; the reflection planes of polyethylene's orthorhombic crystal are labelled on the diffraction rings. **(H)** WAXD Brag 2 θ diffraction measured using film XRD on GP Nano. Gaussian peak fitting was performed on (110) and (200) diffraction peaks. **(I)** DSC thermograms of GP Nano prior and after thermal annealing (145 ℃).*

**Tensile Property Characterization**

The tensile properties of the free-standing GP Nano were probed by in-plane stress-strain and indentation-displacement tests on an advanced rheometric expansion system (ARES), as described in the Supplement (see **Supplement Materials**).

The in-plane stress-strain tests were conducted on rectangular strips that were cut with surgical scalpels containing triangular blades. A video recording of the stress-strain test on film samples with dimensions of $10\ mm \times 15\ mm \times 39\ nm$ at a constant extension rate of $0.001\ s^{-1}$ is shown in **Supplementary Materials**, **movie S1**, in which the sample was extended to a limiting extension strain of 5% along the 2$^{nd}$ planar extension direction of the film during film preparation followed by a load reversal. The stress-strain data during extension are congruent with the measurements conducted on film samples with dimensions of $5\ mm \times 8\ mm \times 59.4\ nm$, **Supplementary Materials, fig. S7**. Thus, the ultimate stress-strain data were measured on the smaller samples ($5\ mm \times 8\ mm \times 59.4\ nm$) all data are measured along the 2$^{nd}$ planar extension direction of the film. **Fig. 3A** depicts the stress-strain responses of the GP Nano at a Hencky strain rate of $0.001\ 1/s$ under ambient conditions. All stress-strain curves are congruent with similar slopes. In addition to exhibiting extraordinarily high tensile strengths of 1071 MPa depicted in **Fig. 1D**, the GP Nano also possess an extremely high work of fracture of 196.7 $MJ/m^3$; this is approximately 50 times that of the densified wood recently developed(*25*). It is also interesting to note that the GP Nano had a strong linear viscoelasticity at strains up to 9 % with a Young's modulus of 10.3 $\pm$ 0.5 GPa. The large linear region observed in the film was attributed to the cellular structure in the film, which had higher effective deformation strains than the solid films.

The annealed monolayer 2D film displayed full linear viscoelasticity until fracture, **Fig. 3D**. The Young's moduli of the monolayer 2D films being $E = 5.0 \pm 1.5$ GPa were lower than those of the multilayer films. The enhanced linearity in the 2D film likely resulted from the larger cell dimensions and stronger triangulated connectivities as depicted in **Fig. 2, E to F**. Besides, the lower crystallinity in the monolayer 2D film was perhaps also helpful. We speculate that the amorphous taut-tie molecules linking the crystalline regions could be more effective in enabling intercrystalline shearing, intercrystalline separation and intercrystalline stack rotation, and forming rubbery structures led to recoverable deformation. On the other hand, deformation of the highly crystalline multilayered GP Nano was likely to include nonlinear deformation of the crystals, such as deformation-induced crystal twinning and stress-induced martensitic deformation. The decrease in the Young's modulus was a direct consequence of a decrease in the crystallinity. As shown in **Fig. 2I**, the annealed film showed a crystallinity of 66.5% which is 12% lower than that in the multilayered GP Nano (the control). To our knowledge, this is the first experimental finding of full linear viscoelasticity in polyethylene. Creep in polymers, especially in polyethylene, has been a long-standing issue yet unresolved.

To the best of our knowledge, the in-plane tensile tests were the first direct measurements of cellular polymer nanofilms due to the hand manipulability of the GP Nano. The maximum load measured is typically approximately 0.3 N (30 gf) for 5 mm wide films, and these load levels are measurable using standard tensile testers. Until now, the best literature direct in-plane data were obtained on solid polystyrene nanofilms using customized atomic force microscopy load sensors with samples floating the surface of water(*11*). The tensile properties in most literature reports were inferred from out-of-plane indentation measurements (*13*, *39*). To align with the literature findings and quantify the flexibility and shape recoverability of the GP Nano under lateral dimension constraints, we conducted indentation-displacement measurements on the GP Nano, and the results are presented below.

The force-indentation responses of the freestanding GP Nano were probed by indenting the centre of a free-standing circular film (diameter = 35 mm, thicknesses=43.1 nm and 48.0 nm) with a frictionless glass ball, as illustrated in **Supplementary Materials**, **fig. S8**. We used a glass ball because its stiffness (~70 GPa) was significantly larger than that of the film (~10 GPa). The 17-mm diameter glass ball was attached to an axial fixture with a 10 mm diameter under the ARES. A circular film sample with a free-standing diameter of 35 mm was attached at its perimeter to the grounded surfaces of a Millipore borosilicate glass funnel via van der Waals attractions. Indentation tests were performed at a constant displacement rate followed by load reversal. This cycle was repeated several times, each at different limiting displacements until film fracture, and at the end of each cycle, there was a 15 min relaxation period before the next cycle was started. While hysteresis was observed, the force-displacement measurements were highly reproducible. The data sets at different limiting displacements were congruent, indicating linear viscoelasticity (**Fig. 3, C and D**). The ultimate flexibility of the film, represented by the maximum recoverable central deflection, was approximately 8 mm, which was approximately 185,000 times the solid film thickness. The maximum rupture force was 1.06 N, equivalent to 2.19 million times the film's weight!

Based on the linear viscoelastic responses for the GP Nano and assuming no pre-stretching, we modelled the indentation force-displacement responses of the GP Nano with elastic membrane theory following the approximate solution by Begley and Martin(*40*):

$$\frac{P}{EhR} = \frac{9\pi}{16}\left(\frac{R}{a}\right)^{9/4}\left(\frac{d}{R}\right)^3 \quad (2)$$

where P is the axial load, $E$ is Young's modulus, $h$ is the film thickness, $R$ is the glass ball radius, $a$ is the radius of the span, and d is the central displacement of the film. The model predictions from **Equation (2)** were superposed on the indentation load-deflection plots in **Fig. 3, C and D**. The Young's moduli that gave the best fit to the data were 8.9 GPa for the control multilayer film (solid thickness = 43.1 nm), and 7.3 GPa for the annealed monolayer 2D film (solid thickness = 48.0 nm). Noting that **Equation (2)** was derived for elastic membranes and ignored effects from the Poisson's ratio, we consider the agreement between its prediction and the measured data depicted in **Fig. 3 C and D** excellent. Furthermore, these tensile moduli are quite close to the values measured by the in-plane stress-strain curves, especially on the monolayer 2D film. This was expected because the monolayer 2D film demonstrated linear viscoelasticity at all extension strains that were tested, satisfying the linearity assumption for **Equation (2)**.

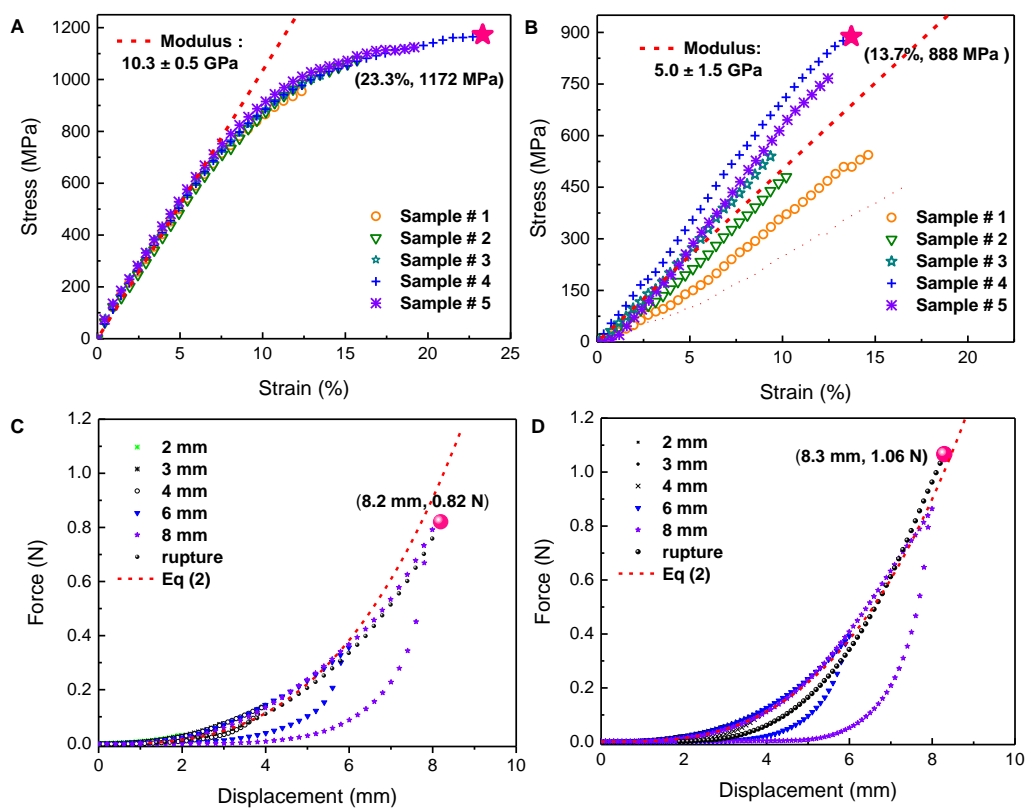

*Fig. 3. Tensile property characterizations of free-standing GP Nano. (A), (B)* In-plane norminal stress strain responses of GP Nano measured on film samples (5 mm × 8 mm) at a constant strain rate of 0.001/s and at the ambient pressure and temperatures prior and after thermal fusional annealing. Five different samples with the same thickness were measured with the extension axis parallel to the 2$^{nd}$ extension directions. The lines represent linear fits to the initial slopes (at strains up to 3%), and the corresponding slopes (Young's moduli) are depicted in the legends. *(C), (D)* Indentation-displacement responses of GP Nano measured on 35-mm diameter films prior to *(3C)* and after partial fusion annealing *(3D)* (see **fig. S8** for the testing configuration). Loop tests were performed on the film samples at limiting displacements of 2 mm to 8 mm. The sample was finally indented to rupture after the loop test. The continuous dashed red lines were calculated using **Equation (2)** derived on elastic membranes.

The unique combination of topological, optical, and mechanical properties in GP Nano is expected to lead to disruptive new technologies(*41*). Herein, we demonstrate its application as a respirator filtration membrane in a transparent face covering.

**Application in Transparent Air Respirators**

Herein, we demonstrate that GP Nano satisfies the standard requirements(*18*) for use as a respiratory face covering in terms of breathability and filtration efficiency tests. A prototype of a respiratory transparent face covering comprising GP Nano is depicted in **Fig. 4A and movie S2**, in which an ultralight GP Nano film with an area density of 0.024 g/m$^2$ and an effective freestanding area of 65.6 cm$^2$ was suspended on a hollow poly (ethylene terephthalate) (PET) scaffold. The freestanding area was designed to ensure the required air flow rate of 85 litres/min at transmembrane pressures below the limiting value of 350 Pa (*18*). This property is a measure of the comfort level; a large flow rate at a small transmembrane pressure is desired for good comfort. The air flux at different transmembrane pressure drops was measured over a freestanding 2-cm diameter sample on a customized air permeation apparatus (**see Supplementary Materials, fig. S9**), **Fig. 4B**. Apparently, the air flow through the membrane followed Knudsen diffusion by exhibiting linear proportionality with transmembrane pressure drops. At a volumetric flow rate of 4 L/min, the value was equivalent to 85 L/min based on the total free-standing area of the face mask (65.6 cm$^2$), and the pressure drop was 146 Pa, which is well below the 350 Pa limit.

The particle filtration efficiency test was carried out using the NIOSH NaCl method, the most rigorous testing method for particle filtration (see **Supplementary Materials**, **fig. S10**), and it was observed that the particle removal efficiency of GP Nano was 99.8%. The morphology of the membrane after particle filtration (penetration) test was characterized under SEM, **Fig. 4B.** The micrograph was taken on a freestanding membrane suspended on a TEM copper grade without any conductive coating, **see Supplementary Materials, fig. S3**. It is seen that all nanofibrils not only retained their integrity, they also kept their taut configurations without any apparent fatigue, suggesting that the membrane was mechanically robust and linearly viscoelastic as alluded in the indentation-displacement responses (**Fig. 3C**). Moreover, salt particles with diameters larger than 100 nm seem to be residing on the top surface of the membrane, and those smaller ones are mostly attached to individual fibril surfaces; it is also interesting to note that the triangulated cellular topology can trap particles at the junctions between two connected cell edges. Particle size distribution was measured using Image J and the measured data are plotted as the inset in **Fig. 4C**. The particles show a very broad range of size distribution with the median at about 66.9 nm. The distribution of particle sizes follows closely with the logistics distribution function. We suggest that this might imply that the size growth of salt particles could be akin to population growth in biological systems (*42*). These particle filtration and breathability test results suggested that GP Nano can act as a passive filter by a sieving mechanism.

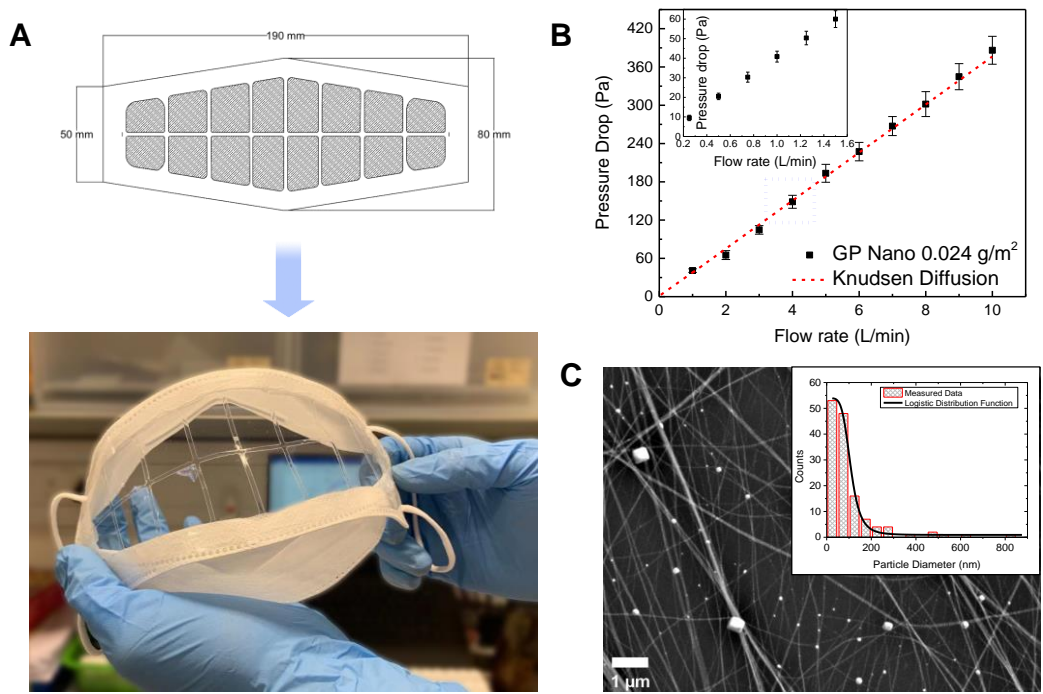

*Fig. 4. Application of GP Nano in transparent face mask and characterizations.* (**A**) *Schematic design and prototype construction.* (**B**) *Pressure drop versus air flowrate measured on a 2-cm diameter*

*freestanding GP Nano (area = 3.14 cm$^2$) at ambient conditions. **(C)** An SEM micrograph of the GP Nano surface depicting the NaCl crystals formed on the membrane surface after filtration tests. The bright dots are NaCl particles. The inset shows the particle size measured using image analysis software Image J; the fitted line is the logistic distribution function.*

**Conclusions**

In summary, a mechanically robust cellular 2D meta-membrane comprising GP Nano with a unique combination of stretch-dominated lattice cells, ultrahigh transparency, ultrahigh mechanical properties, and ultrahigh flexibility is developed. These properties are derived from the microstructures of the cellular film, namely, Delaunay polygonal cells with conformationally anisotropic cell edges. For the first time, direct in-plane tensile properties have been measured on ultralight cellular polymer membranes, and comparisons with customary out-of-plane indentation tests are presented. The mechanical-optical properties of GP Nano with a thickness down to 20 nm, such as the optical transmittance (*34*) and indentation force-displacement relations (*40*), are both analytically describable. Thus, GP Nano is expected to serve as a model platform for further fundamental studies.

For the first time, ultralight materials with area densities of 0.024 g/m$^2$ have been demonstrated to act as a freestanding structural material as a transparent face mask as a result of the unique combination of porosity, transparency and mechanical robustness in GP Nano. The plethora applications for GP Nano can be envisaged as both its cell edges and cell interiors are tuneable. Its application may lead to new technology innovations. If used in lithium-ion batteries, the GP Nano can lead to increase in energy density and battery safety. Because of GP Nano's very low thickness, ultrahigh mechanical strengths and flexibility, its use as a safer separator can allow higher loading of active materials in the battery; its use as a mechanical support for copper by redox plating of copper on its surfaces may lead to at least 80% reduction in copper usage in the electrical current collector of the battery. Besides, the GP Nano can also be used as new skin-conformable and breathable wound patches and biosensors, and new templates for structural evaluations of biomolecules.

**Acknowledgement**

The authors wish to express their sincere thanks to Professor Malcolm R. Mackley of Cambridge University for his insightful comments and suggestions. The authors would also like to acknowledge the support of the Materials Characterization and Fabrication Facility of HKUST. **Funding:** The work is financially supported by the theme-based research scheme of the university grants committee of Hong Kong, Grant No. T23-601/17-R, and the Shenzhen-Hong Kong Innovation Circle, Grant No. SZSTI20EG14.